\documentclass[pdflatex,sn-mathphys-num]{sn-jnl}

\usepackage{graphicx}%
\usepackage{multirow}%
\usepackage{amsmath,amssymb,amsfonts}%
\usepackage{amsthm}%
\usepackage{mathrsfs}%
\usepackage[title]{appendix}%
\usepackage{xcolor}%
\usepackage{textcomp}%
\usepackage{manyfoot}%
\usepackage{booktabs}%
\usepackage{algorithm}%
\usepackage{algorithmicx}%
\usepackage{algpseudocode}%
\usepackage{listings}%

\usepackage{graphicx}	
\usepackage{amsmath}	
\usepackage{eso-pic}
\usepackage{csvsimple}  
\usepackage{dirtytalk}  
\hypersetup{
	colorlinks=true,       
	linkcolor=blue,        
	citecolor=blue,        
	filecolor=magenta,     
	urlcolor=blue         
}
\usepackage{booktabs}   
\usepackage[switch]{lineno}

\begin{document}

\nocite{Einstein1917,Avila2025,Leauthaud2025,Eisenstein2025,Chevallier2001,Linder2003,DESSNKP,DESIDR1,Cole2005,Eisenstein2005,DESIDR2,Planck2020,ACTDR6,SPT3G,Popovic2025,Rubin2026,DESBAO,Hoyt2026,Lodha2025,Chen2025,PDG2025NeutrinoMixing,Linder2007,Calabrese2025,Shajib2025,Brout2022,Rubin2023,Poulin2023,Jedamzik2020,Efstathious2025b,Rosenberg2022,Tristram2024,GarciaQuintero2025,deBelsunce2021,Sailer2025,Jhaveri2025,Giare2024,Munoz2024,Elbers2025a,vincenzi2024,Chung2025,Son2025,Wiseman2026,Murakami2026,Efstathiou2025,vincenzi2025,Cortez2024,Ong2025,Hergt2026,Kids1000,DESY3,HSCY3,DESY6,KidsLegacy,Tsujikawa2026,Wolf2025,Linder2025,Caldwell2025,Wolf2023,Albrecht2006,Schlegel2022,Zhao2024,WST2024,SpecS52025,Prabhu2024,Simons2019,Litebird2023,DESIScience2016,Frohmaier2025,Acevedo2026,ZTFDR22025,Titan2025,Kessler2025,Raghunathan2025,Plavchan2025,Said2025}

\title[$\Lambda$CDM in a courtroom]{$\Lambda$CDM in a courtroom}


\author*[1]{\fnm{Ofer} \sur{Lahav}}\email{o.lahav@ucl.ac.uk}

\author*[1]{\fnm{Paul} \sur{Shah}}\email{paul.shah.19@ucl.ac.uk}

\affil*[1]{\orgdiv{Department of Physics \& Astronomy}, \orgname{University College London}, \orgaddress{\street{Gower Street}, \city{London}, \postcode{WC1E 6BT},\country{UK}}}

\abstract{The cosmological constant $\Lambda$, and its generalization as Dark Energy, makes up roughly 70\% of the present-epoch universe and has been part of the dominant $\Lambda$CDM cosmological paradigm for more than a quarter of a century. It has been put on trial by recent measurements of galaxy clustering, Type Ia supernovae, and the cosmic microwave background. In this Perspective we evaluate the strength of the evidence for a constant $\Lambda$, versus Dynamical Dark Energy (DDE) in the light of recent developments. We also assess what the future holds both for our Universe and our understanding of its most mysterious component.}

\maketitle


\section{Introduction}\label{intro}

Are we witnessing a paradigm shift in cosmology? Although we still do not know the nature of dark matter, an even greater mystery is dark energy (DE). The concept of DE is commonly associated with Einstein’s cosmological constant, $\Lambda$, which he introduced in his 1917 paper \citep{Einstein1917}. In that work, Einstein derived a specific value of $\Lambda$ to obtain a static universe. In contrast, the value of $\Lambda$ inferred today corresponds to a universe undergoing accelerated expansion.

It appears that Einstein regarded $\Lambda$ as a constant of nature, on the left hand side of his equation, alongside the curvature term. Today we might go further than Einstein and propose other additional terms, known as Modified Gravity. If instead we move $\Lambda$ to the right hand side of his equation with the source term, we can view it as an energy density comprising about 70\% of the current Universe, alongside 25\% dark matter and 5\% baryonic (ordinary) matter. We then interpret dark energy as being caused by some unknown microphysics in the Universe, and it becomes possible that this will have some dynamics and change over time. 

The intriguing possibility of a Dynamical Dark Energy (DDE) has led to numerous interpretation and commentary articles, including two Perspective articles and a Q\&A commentary in this journal \citep{Avila2025, Leauthaud2025, Eisenstein2025}. Given the fast developments in the field, within the large collaborations, as well in small teams, we feel it is timely to have a \say{health check} of $\Lambda$CDM and to clarify underlying issues.

\section{What has changed since 2024}

Observational data prior to early 2024 seemed consistent with $\Lambda$.  It is conventional to summarize DE properties using an equation of state parameter $w$, which links its pressure to density by 
\begin{equation}
P = w\rho\;,   
\end{equation}
with constant $w=-1$ corresponding to $\Lambda$.
\par
With the physics of dark energy unknown, cosmologists often use an empirical model for $w$ as a function of the scale factor $a =1/(1+z)$, where $z$ is the redshift:
\begin{equation}
\label{eq:w0wa}
w(a) = w_0 + w_a (1-a)\;. 
\end{equation}
These two parameters, although phenomenological, can summarise a wide variety of physical models \citep{Chevallier2001, Linder2003}. 
In early 2024, the Dark Energy Survey (DES) released a sample of about 1500 Type Ia supernovae (SN) which hinted that DE might be evolving, but at a low statistical significance of $2.5\sigma$ when combined with other data \citep{DESSNKP}. 
Shortly afterwards, the Dark Energy Spectroscopic Instrument (DESI) released measurements of patterns in the distribution of 6 million galaxies in the sky \citep{DESIDR1}, called baryon acoustic oscillations (BAO). First discovered in 2005 by the 2dF and Sloan Digital Sky Survey (SDSS) surveys \citep{Cole2005, Eisenstein2005}, these can be used as \say{standard rulers}, anchoring the expansion history to the Cosmic Microwave Background (CMB). DESI DR2 data released in 2025 \citep{DESIDR2}, incorprating DR1,  increased the galaxies used to 14 million \citep[compared to the 2 million from SDSS used in][]{DESSNKP}.
DESI has now collected spectra of 47 million galaxies and quasars. 

Observations of the CMB by the Planck satellite, Atacama Cosmology Telescope (ACT) and South Pole Telescope (SPT) complement these two data sets to fix the parameters of the early Universe (before DE can exert any influence) very precisely \citep{Planck2020, ACTDR6, SPT3G}.
\par 
The combination of CMB+BAO+DES-SN resulted in a $4.2\sigma$ deviation  from $\Lambda$ \citep{DESIDR2}. However, recent updates to SN data have reduced the significance to $3.2 - 3.4\sigma$ (see Table \ref{table1}) \citep{Popovic2025, Rubin2026}. Nevertheless, this still appears convincing evidence that dark energy may evolve. So, is $\Lambda$CDM dead? What will replace it?

\bigskip
\begin{table}
    \centering
    \renewcommand{\arraystretch}{1.5}
    \resizebox{\textwidth}{!}{\begin{tabular}{l|l|c|c|c|l}
        Model & Data & $ w$ or $w_0$ & $w_a$ & $n\sigma$ & Reference \\
        \hline
        $w$ & Planck-CMB+DESI-DR2-BAO+DES-SN & $-0.971 \pm 0.021$ & - & - & Table V, \citet{DESIDR2} \\
        $w$ & Y6-3x2+DESI-DR2-BAO+DES-BAO+DES-Dovekie-SN & $-0.962^{+0.031}_{-0.029}$ & - & - & Table V, \citet{DESY6} \\
        \hline
        $w_0 w_a$ & Planck-CMB+Y3-3x2+SDSS-BAO+DES-SN & $-0.773^{+0.075}_{-0.067}$ & $-0.83^{+0.33}_{-0.42}$ & $2.5\sigma$ & Table 2, \citet{DESSNKP} \\
        $w_0 w_a$ & Planck-CMB+DESI-DR2-BAO+DES-SN & $-0.752 \pm 0.057$ & $-0.86^{+0.23}_{-0.20}$ & $4.2\sigma$ & Table V, \citet{DESIDR2} \\
        $w_0 w_a$ & \bf{Planck-CMB+DESI-DR2-BAO+Union3.1-SN} & $\mathbf{-0.719 \pm 0.084}$ & $\mathbf{-0.95^{+0.29}_{-0.26}}$ & $\mathbf{3.4\sigma}$ & Table 4, \citet{Hoyt2026} \\
        $w_0 w_a$ & \bf{SPA-CMB+DESI-DR2-BAO+DES-Dovekie-SN} & $\mathbf{-0.803 \pm 0.054}$ & $\mathbf{-0.72 \pm 0.21}$ & $\mathbf{3.2\sigma}$ & Table 10, \citet{Popovic2025} \\

    \end{tabular}}
    \bigskip
    \caption{A selection of recent results from the literature for the dark energy equation of state $w$. The first and second rows illustrate that the "coincidence" of a constant $w$ lying close to $-1$ is not solely due to CMB data.  The remaining rows illustrate the results for the $w_0 w_a$ model using various CMB, BAO and SN data and their significance. DES-SN refers to the results of \citet{DESSNKP}, DES-Dovekie-SN to \citet{Popovic2025}. Planck-CMB refers to the combination of primary power spectrum and lensing reconstructions from either PR3 or PR4 pipelines. SPA-CMB refers to the combination of SPT, ACT and Planck data used in \citet{SPT3G}. Y6-3x2 refers to the results of \citet{DESY6}, and Y3-3x2 to the earlier release in \citet{DESY3}. The entries in bold indicate a high relevance to the discussion on the two competing models.}.
    \label{table1}
\end{table}

\section{The case for Dynamical Dark Energy }\label{caseagainst}

DESI BAO distances to $z \sim 1$ are $\sim 2\%$ closer than $\Lambda$CDM predictions calibrated to Planck CMB data. DES BAO measurements are consistent with DESI \citep{DESBAO}. Conversely, SN distances from DES-Dovekie are $\sim 2\%$ further away at $z \sim 0.05$, when the SN magnitude is calibrated to Planck $\Lambda$CDM \citep{Popovic2025}. While individually each of the CMB, BAO and SN data is well-fit by $\Lambda$CDM, they give different values for the matter density $\Omega_{\rm m}$, indicating $\Lambda$ does not fit all of the data well.

When CMB and DESI BAO data are combined, the evidence for DDE is already $3.1\sigma$ \citep{DESIDR2}. SN data provides insight into the redshift range $z<0.5$ where BAO data starts to be limited by cosmic variance and strengthens the evidence to $3.2 - 3.4\sigma$ \citep{Popovic2025, Hoyt2026}.

The preference for DDE is persistent for alternative parameterisations of $w(z)$ \citep{Lodha2025}, and when substituting any one dataset with an alternative: DESI with DES BAO, DES-Dovekie with Union3.1, Planck with ACT or SPT (but retaining Planck large-scale polarisation data). 

Explanations of the data other DDE have been proposed. \citet{Chen2025} argue that a small, non-zero spatial curvature is preferred by DESI+CMB data. However, this is not supported when SN data is added. 

More evidence against $\Lambda$CDM comes from neutrinos. Laboratory experiments set a bound on the sum of the masses of the individual neutrinos, which is $\Sigma m_\nu > 0.058 $eV (for the mass ordering called normal, as it seems more natural) or $\Sigma m_\nu > 0.098 $eV (inverted mass order) at 95\% confidence \citep{PDG2025NeutrinoMixing}. Neutrino masses affect distances in cosmology by transitioning from a relativistic equation of state ($w=1/3$) to non-relativistic ($w=0$) when their temperature drops below their mass: the higher $\Sigma m_\nu$, the further distances at $z \sim 1$ become. In $\Lambda$CDM, latest CMB+BAO data give $\Sigma m_\nu < 0.048$ eV at a 95\% confidence level \citep{SPT3G}, whereas DDE results relaxes this to $\Sigma m_\nu < 0.129$ eV \citep{DESIDR2}. Hence, if we accept $\Lambda$CDM we have to accept a growing tension with laboratory results.

In summary, the case for DDE is based on the goodness-of-fit of the $w_0 w_a$CDM model vs. $\Lambda$CDM to all data sets, in both Bayesian and Frequentist's approaches. Accepting DDE also relaxes tension with laboratory results on neutrinos.

\section{The case for $\Lambda$CDM}\label{sec:casefor}

By Occam's razor argument, $\Lambda$CDM is simpler than DDE, whose nature must extend beyond the phenomenological parameters $w_0, w_a$. One cross-check is to set the evolution $w_a = 0$, and look for deviation of $w$ from $-1$. If dark energy is dynamical, why should its average value lie close to $-1$? But that is precisely what we do find: see Table \ref{table1}. 

It has been argued by \citet{Linder2007}
that if the CMB angular power spectrum observations (corresponding to $z \approx 1000$) are consistent with $\Lambda$CDM (as tested in \citet{Calabrese2025} for a variety of extensions), the low redshift ($z \sim 0.4$) probes will automatically deliver $w \approx -1$, unless  $w(z)$ is strongly redshift dependent. This is because the distance to the CMB is known regardless of the post-CMB behaviour of dark energy. The information in the CMB power spectrum and standard plasma physics allows us to calculate the absolute size of the fluctuations which later evolve into BAO. As the angular size on the sky is very well measured, the distance to the CMB is then determined. Then, in a loose sense, the average $\langle w \rangle$ for the post-CMB Universe must be close to $-1$ 
to align to the observations. However, even if the CMB is excluded from the analysis, the combination of data from low-reshift probes still gives $w \approx -1$ (e.g. see the second entry in Table \ref{table1}). This argues in favour of $\Lambda$CDM.

\begin{figure}
  \centering
  \includegraphics[width=\textwidth]{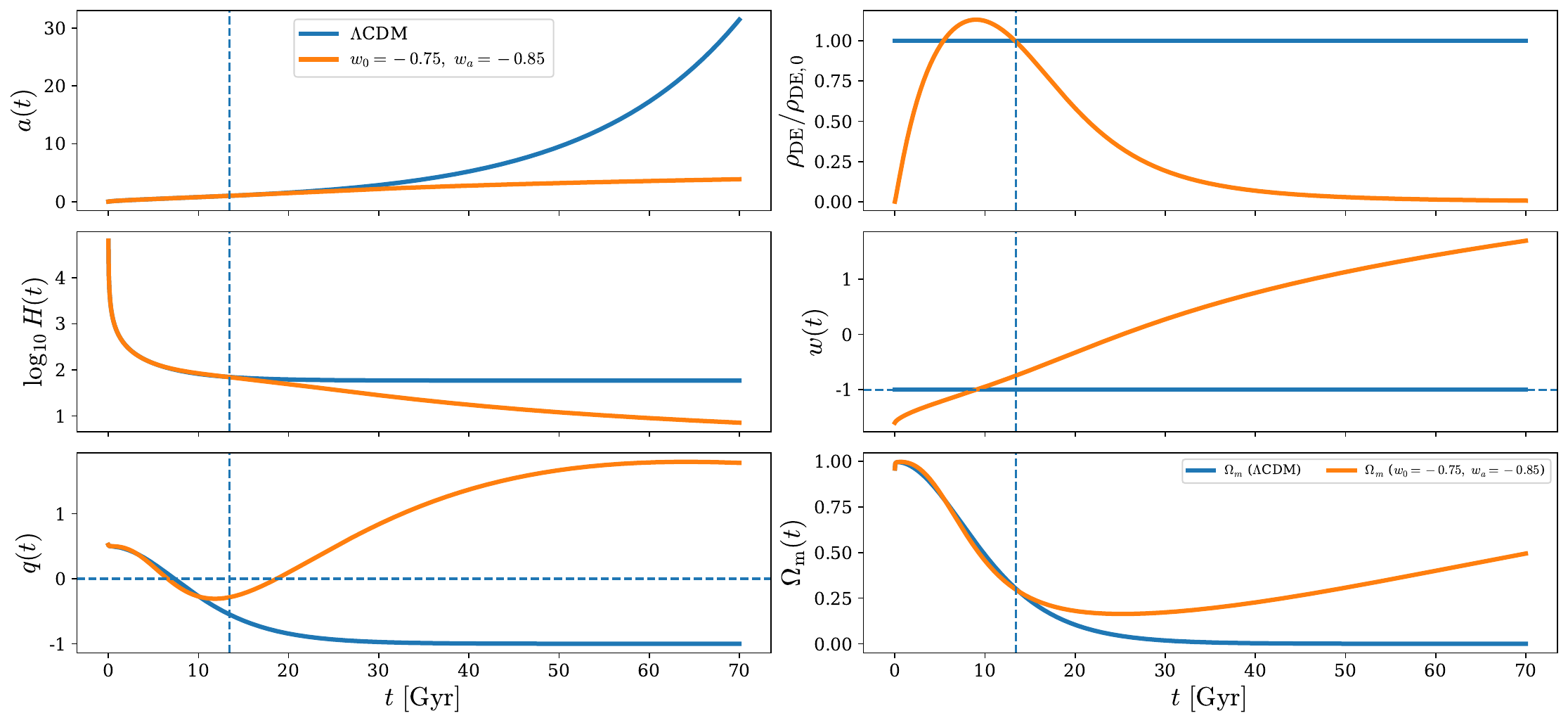}
  \caption{A schematic evolution of six  cosmological quantities with cosmic time, from the big bang to 70 Gyr,
  for rounded values of cosmological parameters. The plots contrast two models: $\Lambda$CDM (blue) and 
  $w_0 w_a$CDM (orange) with indicative values $w_0 = -0.75$ and $w_a= -0.85$, similar to the values given in Table \ref{table1}. Both models assume a flat universe, with $\Omega_m=0.3$ and $H_0= 70$ km/sec/Mpc and negligible radiation. The current age of the universe, 13.5 Gyr in this $\Lambda$CDM model, is marked by a vertical dashed blue line. 
  }
  \label{fig:myplot2}
\end{figure}

A puzzling consequence of this is that if DE evolves and the average $w$ is close to $-1$, then inevitably $w<-1$ at some epoch, called \say{phantom} dark energy. The energy density $\rho_{\rm DE}$ of phantom DE increases with time (Figure \ref{fig:myplot2}), boosting the rate of acceleration of the Universe. If this behaviour persists for long enough, the Universe comes to an abrupt end known as the \say{Big Rip} where DE becomes strong enough to even unbind atoms (fortunately for us, today $w \ge -1$). 

There is growing consensus that physically motivated models for $w(z)$ should replace the phenomenological $(w_0, w_a)$ parameterization in Eqn. \ref{eq:w0wa}. The simplest physical models for DDE, known as quintessence, posit a new scalar field of nature with a non-zero vacuum energy and do not allow $w<-1$ \citep[e.g.][]{Lodha2025, Shajib2025}. While more complex models exist, they introduce \say{non-minimal} terms into the dynamics - which is to say of a form not seen in the current standard model. However, these approaches still require first principles for the setting up the underlying theoretical framework and for the careful choice of priors on the model parameters, and must satisfy all of the probes (including constraints of structure growth and modifications to gravity). Are we willing to accept physics that appears contrived as an alternative to $\Lambda$CDM?

The evidence of the combined probes has been reduced from $3.8 - 4.2\sigma$ to $3.2 - 3.4\sigma$ by improvements to SN standardisation. DES-Dovekie \citep{Popovic2025} and Union3.1 \citep{Rubin2026,Hoyt2026} improve the photometric calibration, population and environmental modelling compared to Pantheon+ \citep{Brout2022}, DES-SN5YR \citep{DESSNKP} and Union3 \citep{Rubin2023} data used by the DESI analysis. The matter density of these new datasets is now more closely aligned, and consistent with the CMB at the $1.0\sigma$ level, although BAO differs by $1.8\sigma$.

Summarising, the case for $\Lambda$CDM rests on two assertions : that there is undiscovered systematic bias in one of SN, CMB or BAO, and the remaining two datasets are a statistical fluke at the level of $1.5-3.1\sigma$. $\Lambda$CDM would also save the need to explain the bizarre phantom behavior of DDE.

\section{Frequently Asked Questions}

\subsection{Trusting the probes}

The process of reducing images to distances requires standardisation methods (such as template fitting), removal of foregrounds (CMB and SN), and merger of data from multiple telescopes (SN). An additional factor is that even when data are independent, analysis pipelines often share features which might introduce common systematics. For example, Pantheon+ and DES share a common pipeline but a small fraction of data, whereas Pantheon+ and Union3 share much data but have independent pipelines. 

\textbf{BAO} measure angles on the sky and redshift differences, converted to distances by a calculation of the sound horizon size. The preference for DDE would be reduced if this size was increased by $\sim 2\%$. This might happen if the Universe before the CMB evolved differently to our expectations: models that change the sound horizon have been mooted to resolve differences in the measurement of the Hubble constant \citep[e.g.][]{Poulin2023, Jedamzik2020}. However, these aim to \textit{reduce} the size of the sound horizon. Moreover, there is currently no evidence from the CMB power spectrum to support them \citep{Calabrese2025}. Interestingly however, the change in the results from DESI DR1 to DR2 was less than might have been expected from the increase in statistical precision alone. This can be traced back to DR2 distances being slightly closer to $\Lambda$CDM calibrated to the CMB compared to DR1 \citep{Efstathious2025b}.

\textbf{CMB} data serve to anchor cold dark matter and baryon densities at high redshift. As explained earlier, while BAO and SN can each be fitted to $\Lambda$CDM, their fits yield different $\Omega_{\rm m}$ to that obtained from the CMB. Rescuing $\Lambda$CDM requires the three $\Omega_{\rm m}$ to agree. Planck data has been re-analysed with different methods of foreground subtraction and likelihood construction \citep{Rosenberg2022, Tristram2024}. Nevertheless some internal anomalies are persistent at the $\sim 2\sigma$ level, which can manifest themselves as preferences for non-zero spatial curvature or anomalous gravitational lensing \citep{Planck2020, Rosenberg2022}. A similar level of difference is visible between the matter densities of ACT and Planck \citep[Figure 37 of][]{ACTDR6}, although SPT is consistent with both. Consequently, the evidence for DDE varies by $\sim 0.3\sigma$ depending on the CMB data combination used \citep{GarciaQuintero2025}.
The matter density from the CMB is partially degenerate with other parameters, including the reionisation optical depth $\tau$. Increasing $\tau$ implies CMB photons are more likely to be scattered before reaching us, damping the amplitude of observed fluctuations. This can be compensated by decreasing the matter density, which influences both the distance to the CMB and also the fluctuation amplitude. $\tau$ is measured at low S/N by Planck large-scale polarisation, it takes values ranging from $0.052 \pm 0.007$ \citep{Rosenberg2022} to $0.063 \pm 0.006$ \citep{deBelsunce2021}. A $\tau$ larger than $0.07$ would decrease the matter density from the CMB enough to make $\Lambda$CDM preferable to DDE \citep{Sailer2025, Jhaveri2025}. Determinations of $\tau$ omitting large-scale Planck data produce higher values of $\tau = 0.078 \pm 0.013$ (\citet{SPT3G}, see also \citet{Giare2024}), and there have also been claims that JWST observations of high-z galaxies also support a high $\tau$ \citep{Munoz2024}. However, these have large uncertainties and a recent compilation of JWST data supports the Planck $\tau$ value \citep{Elbers2025a}.

\textbf{SN} are arguably the most challenging, as their astrophysics is complex. They are rare, so photometry from multiple telescopes must be calibrated to a common system. The interplay of SN environment, progenitor and foregrounds (such as the link between host galaxy mass and dust extinction) is not fully understood, and so observations are empirically adjusted. Adjustments must be made for the bias to see the brighter part of the SN population at deep redshifts due to survey magnitude limits, introducing a population and selection modelling dependency into distances. Estimates of these uncertainties (including evolution with redshift) is included in the data and errors \citep{Popovic2025, Rubin2026, vincenzi2024}. Although there have been recent claims for larger SN evolution \citep{Chung2025, Son2025}, this is not supported by a detailed analysis of the data \citep{Wiseman2026, Murakami2026}. The availability of multiple SN datasets mitigates (but does not eliminate) modelling risks. The Pantheon+ dataset \citep[$2.8\sigma$,][]{Brout2022} is a merger of 20 surveys, with the oldest observation dating from 1980, and models for photometric calibration and host galaxy mass have been revised in the light of new data. DES-SN5YR \citep[$4.2\sigma$,][]{DESSNKP} and its updated calibration DES-Dovekie \citep[$3.2\sigma$,][]{DESSNKP} share the same pipeline as Pantheon+, but combine one high-z survey with six low-z ones, removing older observations and models. The differences between Pantheon+ and DES data reported in \citet{Efstathiou2025} have been shown to be the expected consequence of better host galaxy modelling and selection effects \citep{vincenzi2025}. Nevertheless, the difference between DES-SN5YR and DES-Dovekie ($\sim 2\sigma$ relative to the systematic estimation) serves to illustrate the continuing challenge of cross-survey calibration. While Union3 \citep[$3.8\sigma$,][]{Rubin2023} shares much data with Pantheon+, its Bayesian modelling is an important cross-check of the DES pipeline. Updates to the host galaxy masses and population modelling \citep[$3.2 - 3.4\sigma$, Table 4 of][]{Hoyt2026} bring the results close to DES-Dovekie. In summary, analyses by different teams of (mostly) independent data, using (mostly) independent pipelines are consistent with each other. We discuss prospects to further improve control of systematics in Section \ref{sec:whatsnext}.

\subsection{Interpreting the evidence}
There is a large body of material about how to evaluate the performance of one model relative to another. Frequentist methods consider the probability of data given a model (often by reference to the change in best-fit, $\Delta \chi^2$). Conversely, Bayesian methods consider the probability of the model given the data, incorporating prior belief in model parameters. The Bayesian approach became widely adopted in cosmology in the 1990s, largely displacing the Frequentist methodology. However, Bayesian methods can be criticized for being sensitive to prior assumptions \citep{Cortez2024} and there has recently been a resurgence of Frequentist methods; some DESI analyses combine elements of both approaches \citep{Lodha2025}. 

However, both methods give broadly equivalent conclusions if appropriate thresholds for claiming a discovery are used: the key point is that Frequentist levels of $\sim 3 \sigma$ does not meet the discovery threshold of $5\sigma$. In \citet{Popovic2025}, it was shown that $3.2\sigma$ Frequentist is equivalent to 5:1 odds in favour $w_0 w_a$CDM over $\Lambda$CDM in Bayesian evidence \citep[see also][]{Ong2025,Hergt2026}. Interesting, but not something you would bet the house on! This conclusion is robust to reasonable choices of priors. We note that some $2-3\sigma$ tensions have disappeared with revised data and analyses; a recent example is the tension in the clumpiness parameter $S_8$. It was reported to be larger than $2\sigma$ between weak lensing surveys \citep{Kids1000,DESY3,HSCY3} and the CMB \citep{Planck2020}, but recent results \citep{DESY6, KidsLegacy} find reduced differences. On the other hand, if $5 \sigma$ were to be reached in Frequentist terms, this would likely be equivalent to model odds of better than 1000:1 in the Bayesian calculation. Both methods tell the same story. 

\subsection{Relating dark energy to other cosmological parameters}

An indication in favour of accepting DDE would be to ask: \say{Does it help understand other tensions in cosmological models?}. The answer is no: the tension between the value of the Hubble constant $H_0$ as determined from a local distance ladder, and as determined from CMB+BAO+SN is not substantially changed by DDE \citep{Popovic2025}. The amplitude of matter fluctuations $S_8$ inferred from CMB+BAO remains moderately higher than galaxy survey results \citep{Calabrese2025}. If $\Lambda$CDM is to be replaced, we are looking for more change than DDE.

\subsection{What dark energy may be made of}

The quintessence family of models draw their inspiration from how inflation is believed to have behaved in the early universe. They are therefore thought to be natural candidates to explain DDE. The panel for $\rho_{DE}$ in Figure \ref{fig:myplot2} illustrates that in a DDE universe, if a hypothetical observer had measured the density of dark energy in an otherwise empty box throughout cosmic history, it would have initially increased, peaked at around $z \sim 0.4$ and then declined. This behaviour is difficult to explain in quintessence models \citep{Tsujikawa2026}, although additionally modifying gravity by coupling it to the quintessence field may be viable \citep{Wolf2025}.

Strengthening followed by weakening of the dark energy density could also be explained by interactions between dark energy and dark matter, effectively blending matter with dark energy. However, this would alter the dynamics of gravitational collapse, potentially creating problems for models of galaxy and cluster formation \citep{Linder2025}. Another possibility is that there are two dark energies with different equations of state, but we have confused ourselves by analysing them as if they were one \citep{Caldwell2025}. 

A further problem is that empirical two-parameter descriptions of the expansion history such as $w_0,w_a$ may not be sufficient to resolve the physical model of DDE \citep{Wolf2023}. We are likely to require further constraints from structure growth, gravitational tests, and guidance from theorists about which models are preferred. 

\subsection{The fate of the universe}

Figure \ref{fig:myplot2} shows for each of the two competing models, $\Lambda$CDM and $w_0 w_a$CDM,  the time evolution of the scale factor $a$, the Hubble parameter $H$, the deceleration paramter $q$, dark energy density $\rho_{\text{DE}}$, the equation of state $w$, and the matter density parameter $\Omega_m$ since the Big Bang and into the remote future. These plots are pedagogical, aiming to qualitatively contrast the two models, for approximated observed $\Omega_{\rm m} = 0.3$ and $H_0 = 70$ km sec$^{-1}$ Mpc$^{-1}$in a flat universe for both models, 
 and  $w_0 = -0.75$ and $w_a =-0.85$ for the DDE model. \citep [Error envelopes on similar plots can be found in][]{Lodha2025}. A notable feature is the scale-factor $a(t)$ grows much faster for $\Lambda$CDM, and the Universe continues to expand forever with its expansion rate decreasing down to a constant $H_\infty = H_0 \sqrt {\Omega_{\Lambda} }\approx 59$ km sec$^{-1}$ 
Mpc$^{-1}$. Distant galaxies recede from us and they appear to accelerate away, eventually disappearing from view (gravitationally bound regions, such as the Local Group, do not participate on the global expansion).
By contrast, in Flat $w_0 w_a$CDM the universe also expands forever but in $\sim$6 Gyr from today, matter re-asserts its domination of the Universe and the expansion rate resumes its decline to zero. DE becomes a minor speed bump in cosmic history. 

While both models predict Universe will expand forever, some hypothetical scenarios may instead undergo a future recollapse. One example is a Flat $\Lambda$CDM model with a negative cosmological constant, (e.g. $\Omega_m=1.3, \Omega_{\Lambda}=-0.3$). A second is a closed universe ($\Omega_k < 0$) combined with a dark-energy component that decays in the future ($w_a <  0$): if dark energy becomes sufficiently weak in the future, it may no longer counteract gravitational attraction, allowing the expansion to halt and the Universe to eventually collapse. However, neither of these two alternatives are preferred by the data, and unless some other physics emerges, the Universe will not re-collapse in a \say{Big Crunch}.

\subsection{What future data will tell us} 
\label{sec:whatsnext}

A metric for the precision of DDE measurements is the Figure of Merit (FoM), defined as $1/\sqrt{\rm{Cov}(w_0, w_a)}$ \citep{Albrecht2006}. The best yet is given by the results in \citet{Popovic2025}, which uses data collected by DESI in the first three of five years of nominal operations, plus the latest CMB and SN datasets. The next two generations of spectroscopic surveys, starting with DESI-Run2 \citep{Schlegel2022} and continuing with the Multiplexed Survey Telescope \citep[MUST,][]{Zhao2024}, the Wide Field Spectroscopic Telescope \citep[WST,][]{WST2024} and SPEC-S5 \citep{SpecS52025} would greatly improve the measurement of BAO at higher redshifts. This helps break degeneracies between late-time matter and dark energy densities. For imaging surveys, it is expected that 'Stage-4' projects (e.g. Rubin LSST) would have FoM that is at least 3 times larger than that of 'Stage-3' projects (e.g. DES).

Regarding the CMB, the SPT and Simons Observatory \citep{Prabhu2024, Simons2019} will tighten constraints on parameters that are degenerate with $\Omega_{\rm m}$, such as the spectral index of primordial perturbations $n_s$ and the number of relativistic particles $N_{\rm eff}$. Furthermore, they will help constrain theories that deviate from $\Lambda$ pre-recombination such as Early Dark Energy \citep[see Figure 7 of][]{Poulin2023}. The Litebird satellite, expected to launch in 2032, will measure $\tau$ to an accuracy of 3 times better than Planck. Litebird will also infer neutrino masses with an expected precision of $\sigma(\Sigma m_{\nu}) \sim 0.012$ eV \citep{Litebird2023}, resolving if $\Lambda$CDM values are in conflict with experiment. According to \citet{DESIScience2016}, the final FoM for DESI + CMB will improve on today's value by a factor 1.5.

However, this is unlikely to be enough to claim a discovery of DDE by exceeding the $5\sigma$ threshold, if the central values do not move. 
Hence, the role of SN is very important. Rubin LSST, in conjunction with spectroscopic followup from the 4MOST-TiDES survey will collect two orders of magnitude more SN than the entirety of observations so far \citep{Frohmaier2025}. This data will be supplemented by the low-z DEBASS \citep{Acevedo2026}, ZTF \citep{ZTFDR22025} and ATLAS \citep{Titan2025} surveys, and the high-z Roman time domain survey \citep{Kessler2025}. Forecasts with an early Rubin LSST dataset give a FoM improvement by a factor of 4 after 3 years of observations \citep{Raghunathan2025}. If central values of $w_0, w_a$ do not move much (a big if) from today's results, it is probable that this will be sufficient to claim a $5\sigma$ detection of DDE. 

Central to this forecast is whether control of SN systematics continues to keep pace with the collection of additional data. There are reasons to be optimistic: the new low-z data will allow construction of large volume-limited datasets to improve the understanding of SN astrophysics and foregrounds. Photometric calibration will be improved by the Landolt satellite \citep{Plavchan2025} and peculiar velocity estimation by the DESI survey \citep{Said2025}. 

\section{Conclusions}\label{conclusions}

In this article, we have presented the case for and against $\Lambda$. In our view, neither side presents a case beyond reasonable doubt, leading us to a legal verdict of \say{Not Proven} (up until 2025, Scottish law allowed for this controversial outcome in addition to Guilty or Not Guilty). We warn against the over-interpretation of frequentist statistics at the level of $\sim 3\sigma$, but nevertheless the inconsistencies of $\Lambda$ are hard to dismiss - if they are real rather than a manifestation of undiscovered systematics. 

In the next few years, further releases of BAO and SN data give grounds for optimism on a definitive answer: does DE evolve or not? One could say the future of dark energy is bright.

\bibliography{sn-bibliography-correct-order}

\section{Author contributions}
The authors are members of the DES, DESI and DESC collaborations, but the views expressed here are their own. Both authors contributed equally to this work. 

\section{Statement of competing interests}
The authors declare no competing interests.

\end{document}